\documentclass[12pt,preprint]{aastex}

\newcommand{\Hm}{\rm{H}^{-}}
\newcommand{\me}{\rm{e^{-}}}
\newcommand{\Hp}{\rm{H}^{+}}
\newcommand{\Dp}{\rm{D}^{+}}
\newcommand{\Hep}{\rm{He}^{+}}
\newcommand{\mH}{\rm{H}}

\newcommand{\mD}{\rm{D}}
\newcommand{\He}{\rm{He}}
\newcommand{\mHt}{\rm{H}_{2}}
\newcommand{\hd}{\rm{HD}}
\newcommand{\mHtp}{\rm{H}_{2}^{+}}
\newcommand{\htp}{\rm{H}_{3}^{+}}
\newcommand{\mC}{\rm{C}}
\newcommand{\Cp}{\rm{C^{+}}}
\newcommand{\ch}{{\rm CH}}
\newcommand{\Cm}{{\rm C^{-}}}
\newcommand{\mO}{\rm{O}}
\newcommand{\Om}{\rm{O^{-}}}
\newcommand{\Op}{\rm{O}^{+}}
\newcommand{\oh}{{\rm OH}}
\newcommand{\hto}{\rm{H_{2}O}}
\newcommand{\co}{{\rm CO}}
\newcommand{\msi}{\rm{Si}}
\newcommand{\sip}{\rm{Si^{+}}}
\newcommand{\sipp}{\rm{Si^{++}}}

\def\simless{\mathbin{\lower 3pt\hbox
   {$\rlap{\raise 5pt\hbox{$\char'074$}}\mathchar"7218$}}}   
\def\simgreat{\mathbin{\lower 3pt\hbox 
   {$\rlap{\raise 5pt\hbox{$\char'076$}}\mathchar"7218$}}}

\slugcomment{to be submitted to ApJ}

\shorttitle{Initial Conditions vs.\ Metallicity Thresholds for Star
Formation}
\shortauthors{Jappsen et al.}

\begin{document}

\title{Star Formation at Very Low Metallicity. V. The greater
  importance of initial conditions compared to metallicity thresholds.}

\author{Anne-Katharina Jappsen\altaffilmark{1}}
\affil{School of Physics and Astronomy, Cardiff University, Cardiff, UK}
\email{jappsena@cardiff.ac.uk}

\author{Mordecai-Mark Mac Low\altaffilmark{2}}
\affil{American Museum of Natural History, New York, NY, USA}
\email{mordecai@amnh.org}

\author{Simon C. O. Glover}
\affil{Institut f\"ur Theoretische Astrophysik, Zentrum f\"ur
  Astronomie der Universit\"at Heidelberg, Heidelberg, Germany}
\email{sglover@ita.uni-heidelberg.de}

\author{Ralf S. Klessen}
\affil{Institut f\"ur Theoretische Astrophysik, Zentrum f\"ur
  Astronomie der Universit\"at Heidelberg, Heidelberg, Germany}
\email{rklessen@ita.uni-heidelberg.de}

\and

\author{Spyridon Kitsionas\altaffilmark{3}}
\affil{Astrophysikalisches Institut Potsdam, Potsdam, Germany}
\email{skitsionas@googlemail.com}

\altaffiltext{1}{Fomerly located at the Canadian Institute for Theoretical Astrophysics, Toronto, ON,
  Canada}
\altaffiltext{2}{Also at the Max-Planck-Institut f\"ur Astronomie and
  the Institut f\"ur Theoretische Astrophysik, Zentrum f\"ur Astronomie
  der Universit\"at Heidelberg, Heidelberg, Germany}
\altaffiltext{3}{Currently located at Hellenic-American Educational Foundation, Psychiko College, P. Psychiko, Greece}

\begin{abstract}

The formation of the first stars out of metal-free gas appears to result in stars
at least an order of magnitude more massive than in the present-day case. We
here consider what controls the transition from a primordial to a modern initial
mass function.  It has been proposed that this occurs when effective metal line
cooling occurs at a metallicity threshold of $Z/Z_{\odot} > 10^{-3.5}$. We
study the influence of low levels of metal enrichment on the cooling and collapse
of initially ionized gas in small protogalactic halos using three-dimensional,
smoothed particle hydrodynamics simulations with particle splitting. Our initial
conditions represent protogalaxies forming within a previously ionized H~{\sc
ii} region that has not yet had time to cool and recombine. These differ
considerably from those used in simulations predicting a metallicity threshold, where the gas was initially cold and only partially ionized. In the
centrally condensed potential that we study here, a wide variety of initial
conditions for the gas yield a monolithic central collapse. Our models show no
fragmentation during collapse to number densities as high as $10^5$~cm$^{-3}$,
for metallicities reaching as high as $10^{-1}\,Z_{\odot}$, far above the threshold suggested by previous work. Rotation allows for
the formation of gravitationally stable gas disks over large fractions of the
local Hubble time. Turbulence slows the growth of the central density slightly,
but both spherically symmetric and turbulent initial conditions collapse and
form a single sink particle. We therefore argue that fragmentation at moderate
density depends on the initial conditions for star formation more than on the
metal abundances present. The actual initial conditions to be considered still
need to be determined in detail by observation and modeling of galaxy formation.
Metal abundance may still drive fragmentation at very high densities due to dust
cooling, perhaps giving an alternative metallicity threshold.

\end{abstract}


\section{Introduction}

Observations of old stellar populations reveal no primordial,
metal-free stars \citep{bc05,fjb07}, although the lowest-metallicity stars have
abundance distributions suggesting that they may have formed from gas
polluted by the ejecta of a single supernova \citep[see e.g.][]{tun07}. 
Models of primordial star
formation appear to have converged on the result that primordial stars
formed with an initial mass function (IMF) very different from modern
stars \citep{omu01,omu03,abn02,bcl02,oshn06,yoha06,mck08}. Rather than
predominantly forming stars with masses similar to or less than the
Sun's, primordial stars seem to have been massive, short-lived,
objects.

This raises the question of what controls the transition from the
primordial IMF to the modern one.  \citet*{bcl01} performed
simulations of the collapse of cold gas in a top-hat potential that
included the metallicity-dependent effects of atomic fine structure
cooling.  In the absence of molecular cooling, they found that
fragmentation suggestive of a modern IMF only set in at metallicities
above a threshold value of $Z_{\rm th} \simeq 10^{-4} Z_\odot$.
However, they noted that the neglect of molecular cooling could be
significant.  \citet{omu05} argued, based on the results of their
detailed one-zone models, that molecular cooling would indeed dominate
the cooling over many orders of magnitude in
density. \citet[][hereafter Paper IV]{jkgm07} presented the results of
three-dimensional collapse simulations that included molecular cooling
modeled using the simplified chemical network described in
\citet[][hereafter Paper I]{gj07} and \citet[hereafter Paper
III]{glo07}.  They found that fragmentation similar to that seen by
\citet{bcl01} occurs in models starting from the same initial conditions,
but with metallicities below the threshold, and indeed even with zero
metallicity.

These results suggest that the initial conditions adopted by
\citet{bcl01} may have determined the result much more than it might have
been appreciated at the time. These initial conditions can now be seen
to represent not necessarily the predominant mode of early
low-metallicity star formation: the gas begins at a redshift of $z =
100$ with its metallicity already set, but with its temperature the same
as that of unenriched gas at that redshift, and far below the virial
temperature of the halo into which it is eventually incorporated.
Furthermore, the dark matter halo itself has no central condensation,
but merely local perturbations on a flat, top-hat, potential.

It is therefore important to consider whether a metallicity threshold
appears in simulations with different initial conditions than those
adopted by \citet{bcl01} and Paper~IV.  In this paper, we examine a
different, still idealized, set of initial conditions, and find no
fragmentation even at metallicities {\em above} the claimed
threshold. We therefore argue that fragmentation at moderate density
depends on the initial conditions for star formation more than on the
metal abundances present.  The actual initial conditions to be
considered still need to be determined in detail by observation and
modeling of galaxy formation.  Metal abundance may still drive
fragmentation at very high densities as suggested by \citet{omu05} and
\citet{clark08}, perhaps giving an alternative metallicity threshold.

In \S\ref{sec:method} we describe the numerical methods and initial
conditions used for our models, while in \S\ref{sec:result} we
describe our results.  In \S\ref{sec:discussion} we compare our
results to earlier work to demonstrate the importance of the
environment and initial conditions to the problem.

\section{Numerical Methods}
\label{sec:method}
\subsection{Algorithms}
To assess the relative influence of initial conditions and metallicity
on the cooling and collapse of gas in small protogalactic halos, we
used numerical simulations. During collapse, gas increases in density
by several orders of magnitude, and so it is best simulated by a
numerical method with a high dynamical range. We therefore chose
smoothed particle hydrodynamics (SPH). Overviews of the method, its
numerical implementation, and some of its applications are given in
reviews by \citet{ben90} and \citet{mon92}. We use the parallel SPH code
GADGET version 1.1 \citep{syw01} for our simulations. 

SPH is a Lagrangian method for simulating astrophysical flows, in
which the fluid is represented by an ensemble of particles, with flow
quantities at a particular point obtained by averaging over an
appropriate subset of neighboring SPH particles.  The mass resolution
of a simulation is approximately
\begin{equation}
M_{\mathrm{res}} = 2 N_{\mathrm{neigh}} m_{\mathrm{p}}
\end{equation}
where $N_{\mathrm{neigh}}$ is the number of particles within a given
SPH smoothing kernel and $m_{\mathrm{p}}$ the mass of a single gas
particle. To avoid numerical fragmentation the Jeans mass
$M_{\mathrm{J}}$ has to be resolved \citep{BAT97}: $M_{\mathrm{res}} <
M_{\mathrm{J}}$.  In order to achieve a higher mass resolution, we
refine the mass of the gas particles in regions approaching the Jeans
criterion using the method of \citet{KIT02}. For details on the
implementation of the method see Appendix~\ref{appendix}. 

Particle timesteps are constrained by the Courant-Friedrich-Lewy
condition that signals cross no more than a fraction of a resolution
element per timestep, so they grow increasingly short as the
resolution improves during collapse of high density regions.
Replacing dense cores with artificial sink particles therefore leads
to a considerable increase in computational speed, allowing the
dynamical evolution of the lower-density gas to be followed over
multiple free-fall times. In the runs presented here we implement sink
particles according to the prescription of \citet{BAT95}.  Sink
particles are introduced in regions where the density rises above
$1.25 \times 10^5\,\mathrm{cm^{-3}}$, and accrete gas within a radius
of $0.1\,\mathrm{pc}$. On every timestep, any gas within this
accretion radius that is gravitationally bound to the sink particle is
accreted by it.  The design and implementation of our sink particle
algorithm is discussed in more detail in \citet{JAP05}.
 
\subsection{Chemistry and Cooling}
\label{subsec:chemcool}

\subsubsection{Chemical Model}
The chemical model used in our simulations is the same as that used in paper IV, 
and is discussed in more detail in papers I \& III. It is designed to accurately 
follow the major atomic and molecular coolants of the gas.
Provided that carbon and oxygen are amongst
the most abundant metals, the major coolants will be largely the same as in
local atomic and molecular gas, namely $\mHt$, HD, C, $\Cp$, O, Si, $\sip$, CO,
OH and ${\rm H_{2}O}$ \citep{omu05}. We therefore follow the abundances of
these ten species, together with an additional 29 species that play important
roles in determining the abundances of one or more of these coolants. A full
list of the chemical species included is given in Table~\ref{tab:species}.
The chemical network presented in papers I and III contains 189 collisional
gas-phase reactions between these 39 species, as well as grain surface
reactions, reactions involving the photoionization or photodissociation of
chemical species by ultraviolet radiation, and reactions involving cosmic
rays. For simplicity, in the simulations presented in this paper we do not include
the effects of dust, UV radiation or cosmic rays, and so use a simplified
version of the model that contains only the collisional reactions.

To implement this chemical network within GADGET, we make use of operator
splitting. During a given SPH particle time-step, we first compute the new
SPH densities in the standard fashion, and then update the chemical
abundances by solving a coupled set of rate equations of the form
\begin{equation}
\frac{{\rm d}n_{i}}{{\rm d} t} = C_{i} - D_{i} n_{i},
\end{equation}
where $n_{i}$ is the number density of species $i$, and $C_{i}$ and
$D_{i}$ are chemical creation and destruction terms that generally
depend on the temperature $T$ and the chemical abundances of the
other reactants in the system. In our current implementation, we solve
rate equations for the abundances of only 18 of our 39 species.
The abundances of a further 14 species
that have rapid creation and destruction timescales are determined
under the assumption that chemical equilibrium applies, while the
abundances of the final seven species are determined through the
use of conservation laws for charge and elemental abundance. For
each chemical species in our model, the method of solution used is
summarized in Table~\ref{tab:species}. To ensure numerical stability,
we solve our set of chemical rate equations implicitly, using the DVODE
solver \citep{bbh89}, together with an implicit equation for the specific
internal energy of the particle (see \S\ref{cooling} below).

\subsubsection{Cooling Function}
\label{cooling}
The thermal evolution of the gas in our simulations is modelled using a
cooling function that includes the effects of atomic fine structure cooling
from $\mC$, $\Cp$, O, Si and $\sip$, rotational and vibrational cooling
from $\mHt$, HD, CO and ${\rm H_{2}O}$, Lyman-$\alpha$ cooling,
Compton cooling, and $\Hp$ recombination cooling, as well as a number
of other processes of lesser importance. A full list of the processes included
is given in Table~\ref{cool_model}; further details can be found in papers I and III.
To allow for the effects of cosmic microwave background (CMB) heating of the gas, we adopt in our simulations
a modified cooling rate of the form
\begin{equation}
\Lambda = \Lambda(T) - \Lambda(T_{\rm CMB}),
\end{equation}
where $\Lambda(T)$ is the cooling rate of the gas per unit volume for gas
temperature $T$ and $\Lambda(T_{\rm CMB})$ is the cooling rate when
$T = T_{\rm CMB}$. This modification has a negligible effect when
$T \gg T_{\rm CMB}$, but prevents the gas from cooling beneath the floor
set by the CMB temperature, except by adiabatic expansion.

To treat radiative cooling within the GADGET framework, we use the same
isochoric approximation as \citet{syw01}. During a given particle time-step,
we first compute $\dot{u}_{\rm ad}$, the rate of change of the internal energy
due to adiabatic gas physics. We then solve an implicit equation for the new
internal energy:
\begin{equation}
u^{n+1} = u^{n} + \dot{u}_{\rm ad} \Delta t - \frac{\Lambda \left[\rho^{n}, u^{n+1}\right]
\Delta t}{\rho^{n}},
\end{equation}
where $u^{n}$ and $u^{n+1}$ are the internal energy per unit mass at time
$t ^n$ and $t^{n+1}$ respectively, $\rho^{n}$ is the gas density at time $t^{n}$.
This implicit equation is solved simultaneously with the chemical rate equations
using the DVODE solver.

We ensure that the internal energy of the SPH particle does not change by
a large amount during a single timestep by constraining the timestep:
\begin{equation}
\Delta t \leq \eta_{\rm c} \times {\rm min} \left( \frac{u^{n}}{\dot{u}_{\rm ad}},
\frac{\rho^{n} u^{n}}{\Lambda} \right),
\end{equation}
where $\eta_{\rm c} = 0.01$.
This limitation is necessary to ensure that the estimates constructed by
GADGET of the internal energy and thermal pressure at intermediate
points during the timestep are accurate, and hence that the pressure
forces acting on particles with shorter timesteps are also computed
accurately.

\subsection{Initial Conditions}
\subsubsection{Initial Temperature and Density Distribution}
\label{subsec:ICs}

Our initial conditions are based on those used in \citet[][hereafter
Paper II]{jgkm07}, although collapse is followed to much higher
density in the current work. We study protogalaxies forming from gas with varying degree of metallicity. We assume the gas is initially fully ionized, with no molecules present, and with temperature $T_{\rm g} = 10^4$~K.
These initial conditions are intended to represent a fossil H~{\sc ii} region \citep{oh03} polluted by supernovae from the ionizing object.  Because of the higher
mixing rates expected in gas with higher sound speed, such warm gas
will have metals mixed into it far earlier than cold primordial gas
in the same region.  Fossil H~{\sc ii} regions subject to cooling and
collapse will be common as the characteristic lifetimes of ionizing
sources are significantly shorter than the Hubble time even at high
redshifts. The initial temperature of gas in such a fossil H~{\sc ii} region
should, strictly speaking, depend on various factors, such as
the spectrum and strength of the ionizing source and the density
of the gas. However, since the gas cools rapidly through Ly-$\alpha$
cooling at the start of our simulations and since the temperature
at which Ly-$\alpha$ cooling becomes ineffective does not depend on
the initial temperature of the gas, we do not expect our results
to be sensitive to small changes in the initial value of $T_{\rm g}$.
Because of the cooling, the gas collapses into any low-mass dark matter halos present. 

We model one such halo by using a fixed background potential.  The
potential is spherically symmetric with a density profile
\citep{nfw97}
\begin{equation}
\rho_{\mathrm{dm}}(r)=
\frac{\delta_{\mathrm{c}}\rho_{\mathrm{crit}}}{r/r_{\mathrm{s}}
(1+r/r_\mathrm{s})^2},
\label{nfw-profile}
\end{equation}
where $r_{\mathrm{s}}$ is a scale radius, $\delta_{\mathrm{c}}$ is a
characteristic (dimensionless) density and
$\rho_{\mathrm{crit}}=3H^2/8\pi G$ is the critical density for
closure. Note that we use a value for the Hubble constant of $H =
72\,\mathrm{km}\,\mathrm{s}^{-1}\,\mathrm{Mpc}^{-1}$
\citep{spe03}. Following \citet{nfw97} we calculate the characteristic
density and the scale radius from a given redshift and dark halo
mass. We truncate the halo at the radius at which the value of
$\rho_{\mathrm{dm}}$ given by Equation~\ref{nfw-profile} equals the
cosmological background density at the beginning of the simulation.

In all of the runs presented here, the halo mass is $7.8 \times
10^{5}\,{M_{\odot}}$ and the initial redshift is $z=25$. We truncate
the halo at a radius~$r_{\mathrm{t}} = 0.49\,\mathrm{kpc}$. The virial
temperature of the resulting halo is $1900\,\mathrm{K}$ and the virial
radius is $0.1\,\mathrm{kpc}$. In physical units, the scale radius of
the halo is $29\,\mathrm{pc}$ and the full computational volume is a
box of side length $1\,\mathrm{kpc}$. 

We use periodic boundary conditions for the
hydrodynamic part of the force calculations to keep the gas bound within the
computational volume. The self-gravity of the gas and
the gravitational force exerted by the dark matter potential are not
calculated periodically since we assume that other dark matter halos and their
gas content are distant enough to neglect their gravitational influence.

We begin our simulations with a uniform distribution of gas with an
initial density $\rho_{\mathrm{g}}$, taken to be equal to the
cosmological background density, and then allow the gas to relax
isothermally until it reaches hydrostatic equilibrium. This
initial phase of the simulation is merely a convenient way to generate
the appropriate initial conditions for the simulation proper, and so
we do not include the effects of chemical evolution or cooling during
this phase.

The mass of gas present in our simulation was taken to be a fraction
$\Omega_{\mathrm{b}}/\Omega_{\mathrm{dm}}$ of the total mass of dark
matter, where the dark matter density
$\Omega_{\mathrm{dm}}=\Omega_{\mathrm{m}}-\Omega_{\mathrm{b}}$, and
where $\Omega_{\mathrm{b}}$ is the baryon density and
$\Omega_{\mathrm{m}}$ is the matter density. We take values for the
cosmological parameters from \citet{spe03} of
$\Omega_{\mathrm{b}}=0.047$ and $\Omega_{\mathrm{m}}=0.29$, giving us
a total gas mass of
$M_{\mathrm{g}}=0.19\,M_{\mathrm{dm}}$. $M_{\mathrm{dm}}$ is the sum
of the halo mass and the mass of the dark matter background in the
simulated volume, and has a value $M_{\mathrm{dm}} = 1.84 \times
10^{6}\,{M_{\odot}}$. Therefore, $M_{\mathrm{g}}=3.5 \times
10^{5}\,{M_{\odot}}$. Initially, we use $1.4 \times 10^{5}$ SPH
particles to represent the gas, and so each particle has a mass
$m_{\mathrm{p}} = 2.5\,{M_{\odot}}$. Due to the refinement, the
number of particles rises during the simulation up to $9 \times
10^{6}$, and the SPH particles that represent the gas in the
collapsing region have a mass of $m_{\mathrm{p}} =
0.015\,{M_{\odot}}$ (see Appendix~\ref{appendix}). Our SPH smoothing kernel encompasses
approximately 50 particles and since we need twice this number in
order to properly resolve gravitationally bound objects \citep{BAT97},
our mass resolution is $M_{\mathrm{res}} \simeq
100\,m_{\mathrm{p}} = 1.5\,{M_{\odot}}$.

In the runs presented here we introduce sink particles with an
accretion radius of $0.1\,\mathrm{pc}$ when the density rises above
$1.25 \times 10^5\,\mathrm{cm^{-3}}$. At this density $M_{\mathrm{J}}$
exceeds $M_{\mathrm{res}}$ as long as $T > 30 \, \rm{K}$. As this is
much smaller than the CMB temperature at $z=25$, we always satisfy the
\citet{BAT97} resolution criterion by a comfortable margin.

We follow the simulations until collapse occurs and a sink forms or as
close as possible to a Hubble time within reasonable computing
time. The Hubble time at a redshift of $z=25$ is $\sim
100\,\mathrm{Myr}$. After more than a Hubble time our assumption of
an isolated dark matter halo will no longer be valid, since most halos
will have undergone at least one major merger or interaction by this
time. The dynamical time within the halo varies between
$4\,\mathrm{Myr}$ in the center and $450\,\mathrm{Myr}$ in the outer
regions.

\subsubsection{Metallicity}
\label{subsec:metal}

We study both zero metallicity gas and gas that has been enriched to
$10^{-3}\, {Z}_{\odot}$.  A metallicity of $Z=10^{-3}\, {Z}_{\odot}$ is an upper limit derived from QSO absorption-line studies
of the low column density Lyman-$\alpha$ forest at $z \sim 3$
\citep{PET99}. Estimates of the globally-averaged metallicity produced
by the sources responsible for reionization are also typically of the
order of $10^{-3}\, {Z}_{\odot}$ \citep[see e.g.][]{RIC04}.  We
also run a model with a metallicity of $Z=10^{-1}\, {Z}_{\odot}$
to investigate the influence of an even higher metallicity.

In our simulations of metal-enriched gas, we assume that mixing is
efficient and that the metals are spread out uniformly throughout the
computational domain.  We also assume that the relative abundances of
the various metals in the enriched gas are the same as in solar
metallicity gas; given the wide scatter in observational
determinations and theoretical predictions of abundance ratios in very
low metallicity gas, this seems to
us to be the most conservative assumption. However, variations in the
relative abundances of an order of magnitude or less will not
significantly alter our results. We have denoted runs with zero
metallicity with ``Z0'' and runs with metallicity with ``Z-3'' and
``Z-1''.

\subsubsection{Rotation}
\label{subsec:rot}

We further investigate the influence of rotation on the collapse and
possible fragmentation of the gas. Theoretically, we do expect
rotation to become important on scales $r \ll \lambda r_{\rm vir}$,
where $r_{\rm vir}$ is the virial radius of the halo and $\lambda$ is the dimensionless spin parameter, given by
\citep{peeb71}
\begin{equation}
\lambda = \frac{J |E|^{1/2}}{G M^{5/2}},
\end{equation}
where $J$ is the total angular momentum of the halo,
$M$ is the halo mass and $E$ is the total (kinetic
plus potential) energy of the halo. Previous work
has shown that for halos with the range of masses and
redshifts considered in this paper, $\lambda$ has a
lognormal distribution, with mean $\bar{\lambda} =
0.035$ \citep{YOS03}.

In Paper II we showed that rotation has very little effect on the
evolution of the gas at early times but it did appear to affect the
evolution once the density exceeds $n = 100 \: {\rm cm^{-3}}$,
significantly slowing the collapse. Here we investigate if the
rotation results in the formation of a rotationally supported disk.

We perform three simulations in which the initial gas distribution was
given a non-zero angular momentum.  Within the virial radius of the
halo, the gas was placed into rotation with constant angular
velocity. At larger radii, the initial angular velocity decreased
linearly with radius, reaching zero at the truncation radius of the
halo. These runs, which we designate hereafter using ``ROT'', had a
spin parameter $\lambda = 0.05$.

\subsubsection{Turbulence}
\label{subsec:turb}

Finally, we study the influence of low and intermediate levels of
turbulence on the evolution of the gas within the dark matter
halo. Turbulence establishes a network of interacting shocks, where
converging flows and shear generate filaments of high density. We
investigate if the turbulence can provide high density gas that acts
as a seed for fragmentation.  For one run we assume a turbulent energy
of 5\% of the internal energy and denote it with ``TURB1'', while for
two further runs with turbulence we take a value of 10\% and denote it
with ``TURB2''.  We have included turbulence in our version of the
code that is driven uniformly with the method described by
\citet{MAC98} and \citet{MAC99}. We insert energy on scales of the
order of the size of our computational domain, i.e. with wave numbers
$k=1..2$.

\subsubsection{Choice of Parameters}
\label{subsec:choice}
In the choice of parameters, we first restrict ourselves to the
primordial case. This enables us to directly compare our results
with those of Paper IV and to study the influence of differences in
the initial conditions on the dynamical evolution. Since we suspect
that the radial symmetry of our initial conditions in run Z0 may
artificially suppress fragmentation, we investigate the effects of
adding turbulent energy to the gas: a low level of energy in run
Z0-TURB1 and a higher level in run Z0-TURB2.

We next investigate the effects of enriching the gas with a low
level of metallicity, $Z=10^{-3}\, {Z}_{\odot}$. In this case,
we start our investigation with a run which incorporates turbulence
at the TURB2 level; since we find no fragmentation in this case,
we can reasonably assume that we will not find it in runs with less
turbulence, or with the spherical symmetry unbroken. We therefore do not
include the corresponding runs Z-3-TURB1, and Z-3 in our analysis.

Finally, we study the effects of rotation. Here we consider three
different metallicities, the purely primordial case (Z0-ROT),
as well as $Z=10^{-3}\, {Z}_{\odot}$ (Z-3=ROT) and $Z=10^{-1}\,
{Z}_{\odot}$ (Z-1-ROT). We add one run with a high metallicity to
investigate the influence of the metals in the rotating case more
thoroughly.   We do not mix turbulence and rotation in any of the
models to prevent confusion of the respective effects. A complete
list of runs is given in Table~\ref{tab:runs}.

\section{Results}
\label{sec:result}

\begin{figure}
\epsscale{.80}
\plotone{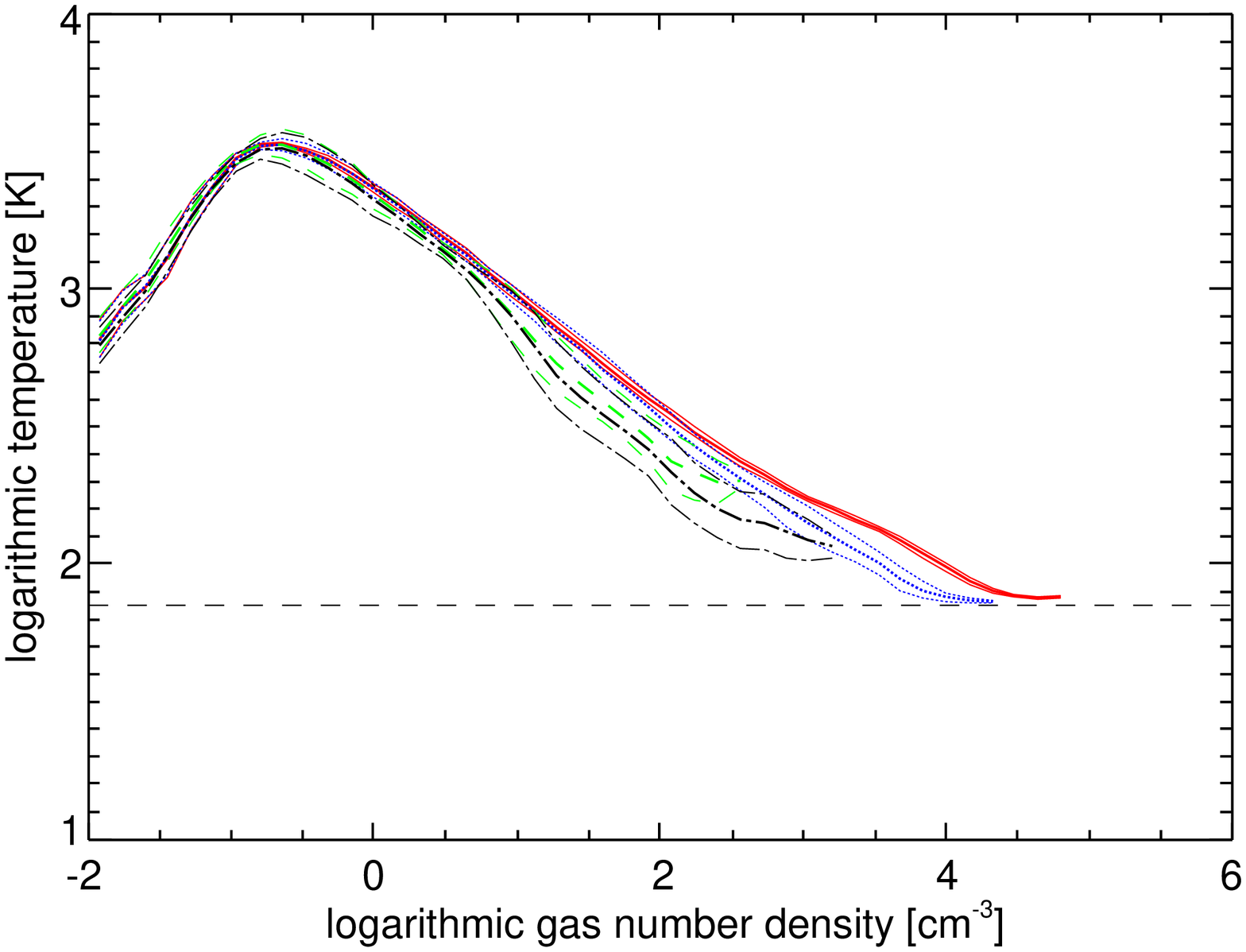}
\caption{Gas temperature vs.\ number density for runs Z0 ({\it red solid
line}), Z0-TURB1 ({\it blue dotted line}), Z0-TURB2 ({\it green dashed line}),
and Z-3-TURB2 ({\it black dot-dashed line}). The thin lines show the
1$\sigma$-deviation around the mean. The time of all plots is $t=52\;
{\rm Myr}$, except for plot Z0-TURB2, which is at $t=50\; {\rm
Myr}$. At this time the gas has cooled down from the initial temperature of $10^4$~K, initially mainly due to Ly-$\alpha$ cooling and later also due to molecular cooling and atomic fine structure cooling if metals are present. We show the CMB temperature with the horizontal dashed
line. Runs Z0 and Z0-TURB1 reach the CMB temperature in $t=52\; {\rm
Myr}$ just from $\mHt$ cooling despite their lack of metals. On the
other hand, run Z-3-TURB2, which has a higher fraction of turbulent
energy, does not reach the CMB temperature in $t=52\; {\rm Myr}$
despite having metallicity Z$=10^{-3}\,Z_{\odot}$.
\label{fig1}}
\end{figure}

\begin{figure}
\epsscale{.80}
\plotone{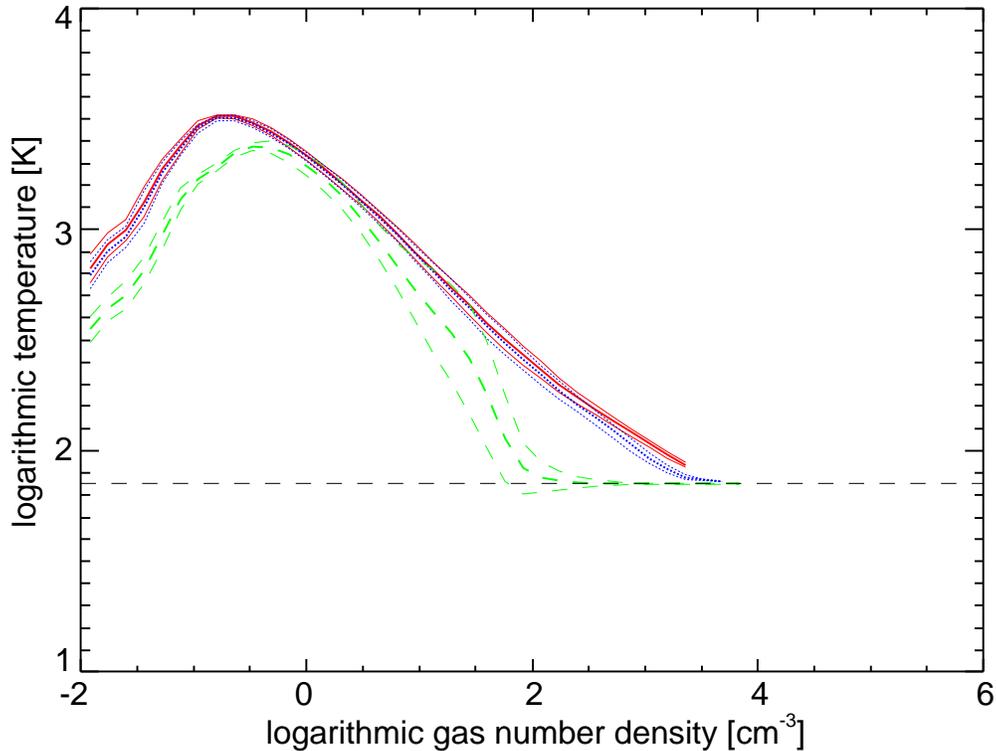}
\caption{ Gas temperature vs.\ number density for runs Z0-ROT ({\it
red solid line}), Z-3-ROT ({\it blue dotted line}), and Z-1-ROT ({\it green dashed
line}). The thin lines show the 1$\sigma$-deviation. The halos with
rotation have a spin parameter of 0.05. The time of the plots is
$t=52\; {\rm Myr}$ as in Figure~\ref{fig1}. We show the CMB temperature with the horizontal
dashed line. The run Z-1-ROT reaches the CMB temperature at a
relatively low density value due to the high content of metals that
contribute effectively to the cooling. The other two runs show only
little difference at higher densities due to the metallicity of run
Z-3-ROT.  \label{fig2}}
\end{figure}

In Figures \ref{fig1} and \ref{fig2} we show the state of the gas after 52 Myr. During this time the gas has cooled down from $10^4$~K, initially mainly due to Ly-$\alpha$ cooling and later due to molecular cooling and atomic fine structure cooling where metals are present. In agreement with the results presented in Paper IV, the inclusion of molecular hydrogen cooling largely erases the difference in
temperature structure between zero and low-metallicity gas (see Figure~\ref{fig1}). Only gas with metallicity as high as $0.1\,Z_{\odot}$ cools at substantially lower density than zero metallicity
gas, as seen in the rotating models shown in Figure~\ref{fig2}.  In
all cases, the gas is able to cool down to the CMB temperature at our
initial redshift $z= 25$.

In Paper~II we found that the centrally concentrated halo profile that we have
chosen to study suppressed fragmentation at number densities below
$500$~cm$^{-3}$ even for metallicities far above the threshold $Z_{\rm
  th}$ suggested by \citet{bcl01}. Our current models show no
fragmentation during collapse to number densities as high as
$10^5$~cm$^{-3}$ for metallicities reaching as high as $10^{-1}\,Z_{\odot}$ in
one rotating case.

We consider several different variations on the initial conditions of
the gas in order to understand the lack of fragmentation that we find.
Model~Z0 is a zero metallicity model that collapses spherically, with
temperatures dropping to the CMB temperature (see Fig.~\ref{fig1}).
In Paper IV, we showed that fragmentation proceeds readily at these
temperatures and densities in a top-hat potential with small
perturbations, even at zero metallicity, but in the centrally
concentrated potential studied here, no fragmentation occurs (top
panel of Fig.~\ref{fig3}). One central sink particle forms and
accretes.

\subsection{Turbulence}

\begin{figure}
\epsscale{.60}
\plotone{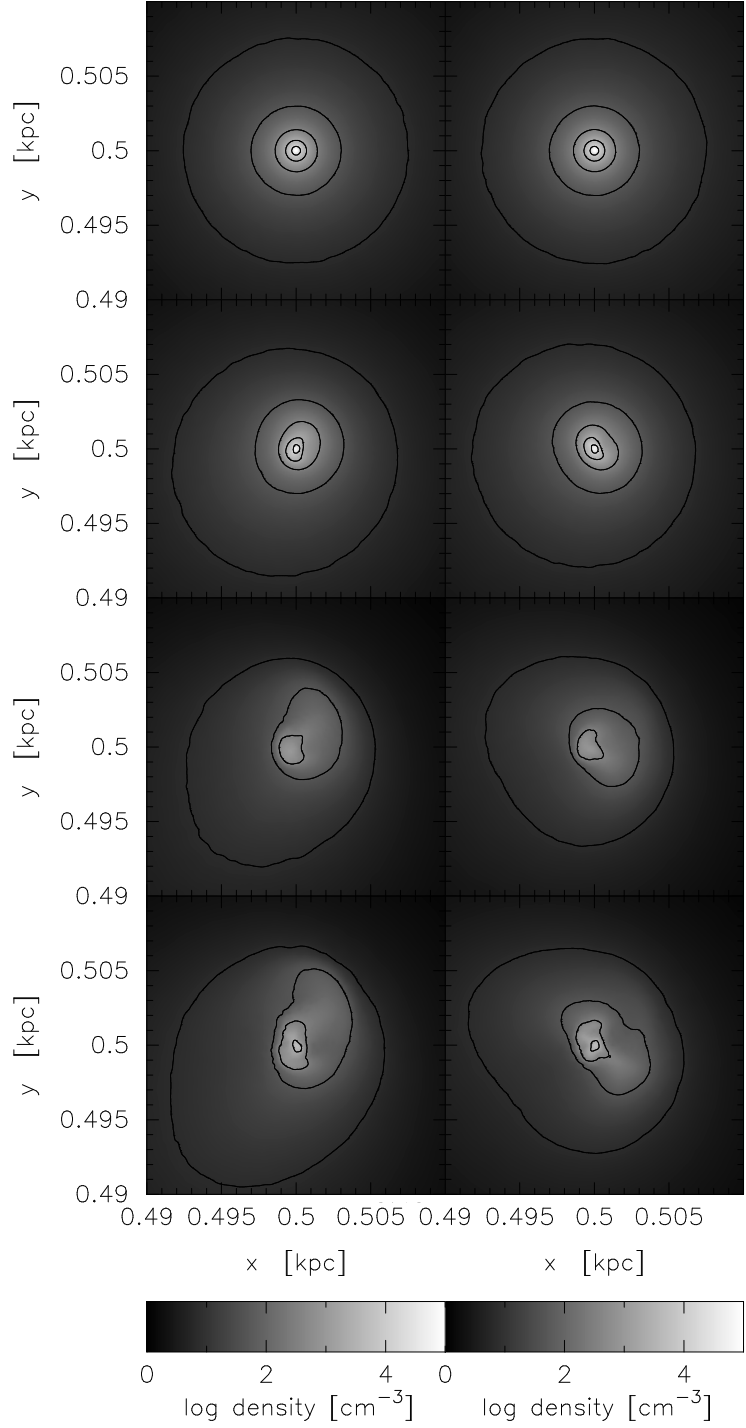}
\caption{{\it From top to bottom}: Cross sections showing number
density for runs Z0, Z0-TURB1, Z0-TURB2, and Z-3-TURB2 at 50 Myr. {\it Left}: x-y plane, cut at $z=0.5$~kpc. {\it
Right}: x-z plane, cut at $y=0.5$~kpc.
\label{fig3}}
\end{figure}

We then consider whether breaking spherical symmetry can lead to
fragmentation.  To do this, we initialize the gas with two different
levels of turbulent kinetic energy (Table~\ref{tab:runs}) in runs
Z0-TURB1 and Z0-TURB2. Converging flows effectively produce randomly
oriented density perturbations, as shown in the middle panels of
Figure~\ref{fig3}.  However, no further gravitational fragmentation
occurs. In the less turbulent TURB1 run, a sink particle forms from
collapsing gas during the run, while even that does not occur in the
more energetic TURB2 runs.  Increasing the metallicity to
$Z=10^{-3}\,Z_{\odot}$ in run Z-3-TURB2 neither markedly increases the
cooling compared to the zero metallicity case (Fig.~\ref{fig1}), nor
produces fragmentation (bottom panel of Fig.~\ref{fig3}).

\subsection{Rotation}

\begin{figure}
\epsscale{0.8}
\plotone{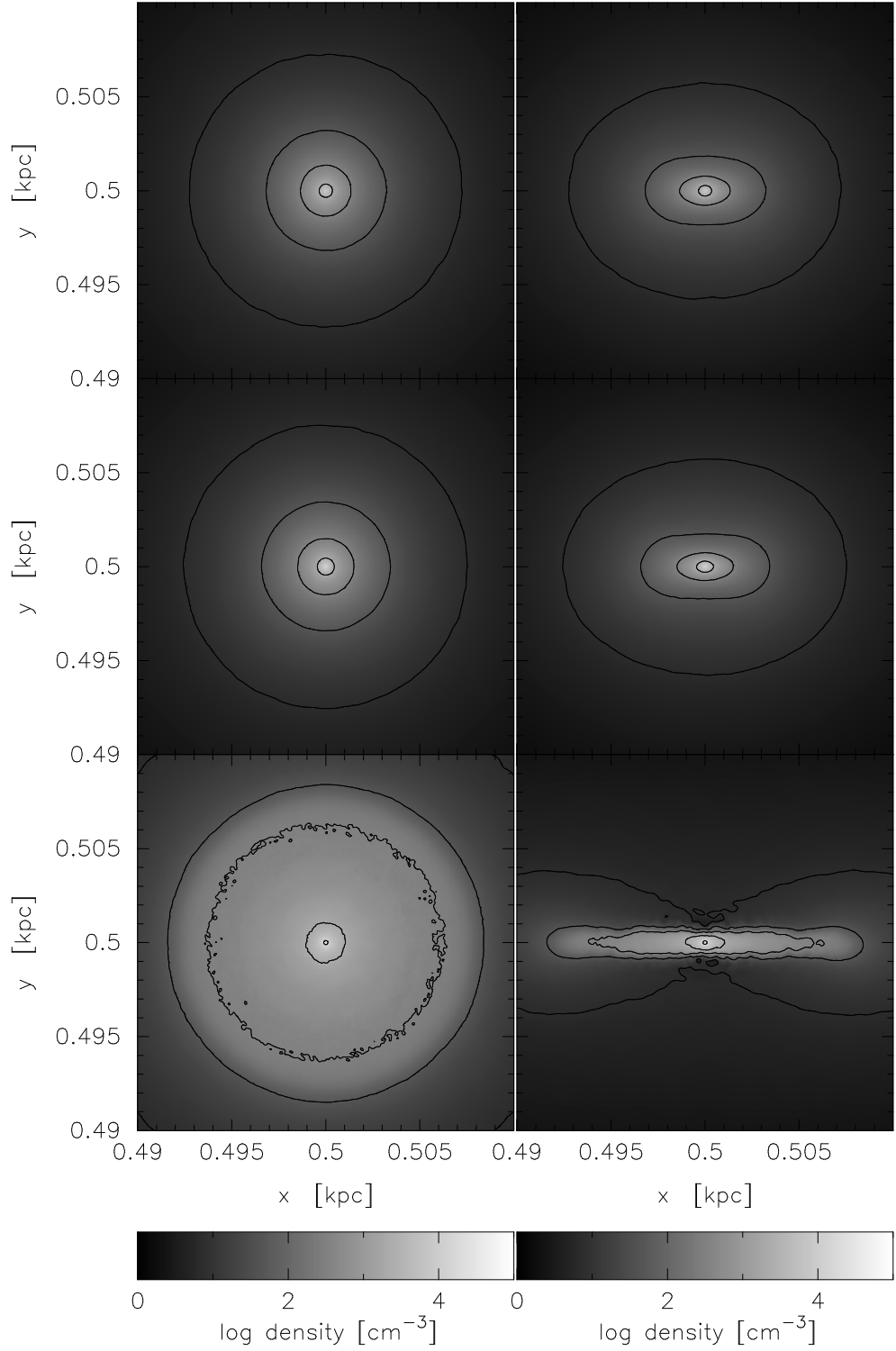}
\caption{{\it From top to bottom}: Cross sections showing number
density for runs Z0-ROT, Z-3-ROT, and Z-1-ROT, at 50 Myr. {\it Left}: x-y plane,
cut at $z=0.5$~kpc. Face-on view. {\it Right}: x-z plane, cut at
$y=0.5$~kpc. Edge-on view. \label{fig4}}
\end{figure}

Up to this point organized rotation has not been introduced into our
models.  Disk formation naturally leads to fragmentation and the
formation of multiple stellar systems in present-day star formation 
\citep[e.g.][]{matsumoto97}, so we investigate whether it
could play a similar role in this situation. In Figure~\ref{fig4} we
show horizontal and vertical number density cross sections of runs in
this parameter study.  We find that thin disk formation proceeds only
in the Z-1-ROT case, with metallicity $Z = 0.1\,Z_{\odot}$. In models
Z0-ROT and Z-3-ROT, with metallicity on either side of the threshold
$Z_{\rm th}$, however, the gas cannot cool strongly enough to form a
thin disk, and merely forms a slightly flattened, pressure-supported
structure. In none of these three models does fragmentation
set in, nor does the central density even grow fast enough for a sink
particle to form within 60~Myr, a significant fraction of a Hubble time at that redshift.

\section{Discussion}
\label{sec:discussion}

The basic thermodynamic behavior shown in Figures~\ref{fig1}
and~\ref{fig2} underlies our conclusion both in Paper~IV and here that the
introduction of metal line cooling does not yield a substantial
difference in star formation behavior once $\mHt$ and HD cooling is also taken
into account.  Recall that the fragmentation threshold found by \citet{bcl01}
occurred in simulations neglecting these coolants.

\begin{figure}
\includegraphics[width=8cm]{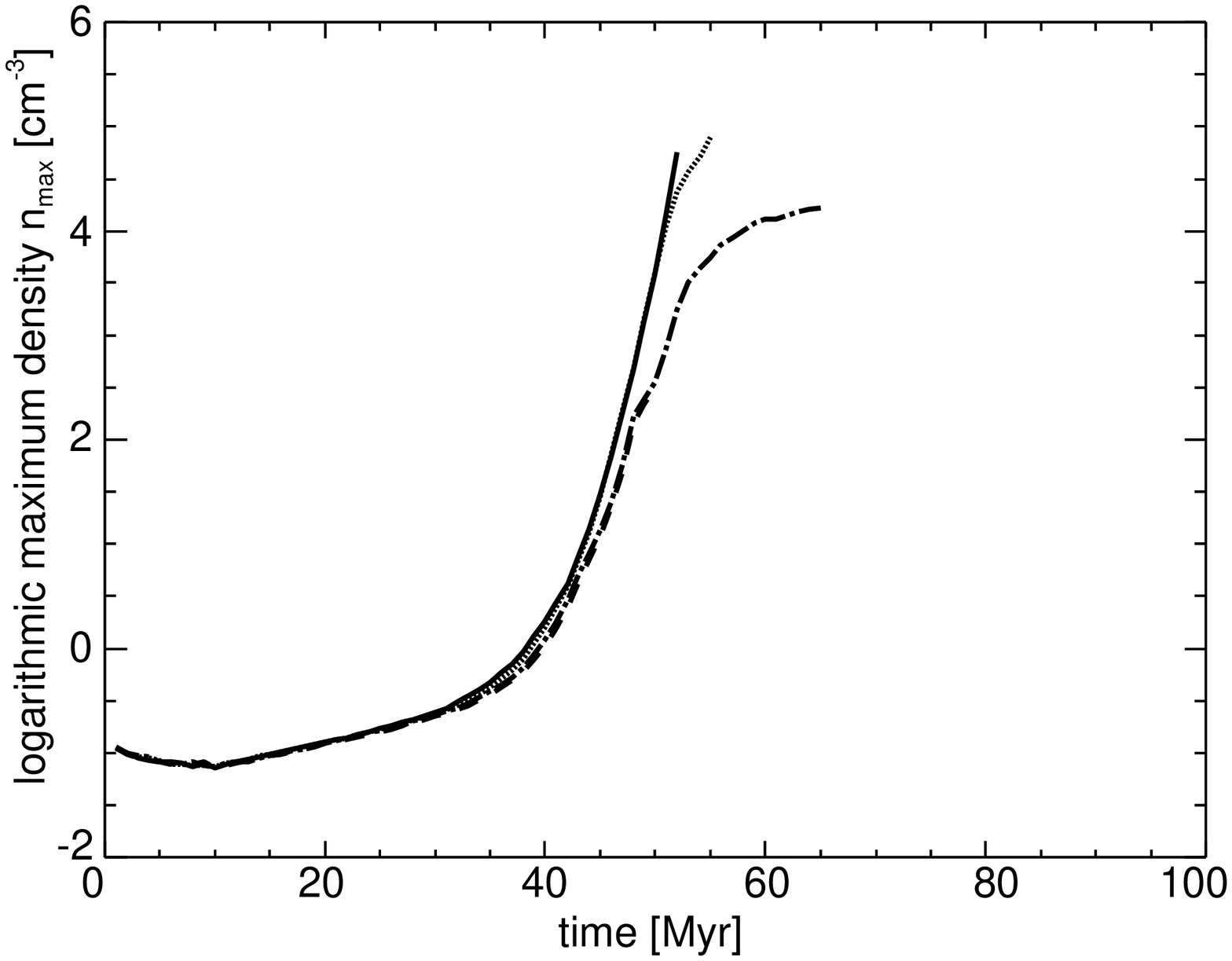}
\includegraphics[width=8.2cm]{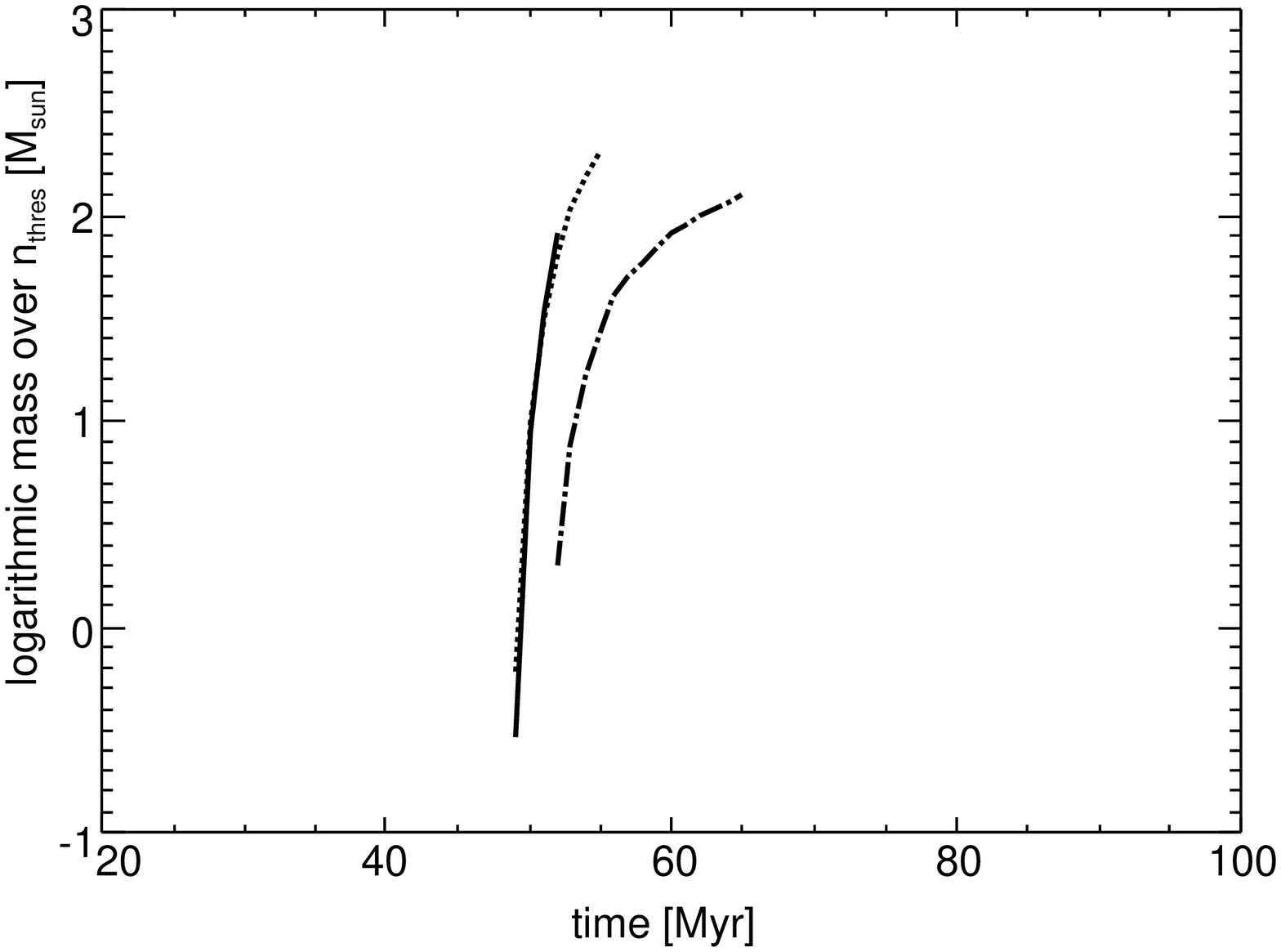}
\caption{{\it Left panel:} Maximum gas number density vs time for runs
  Z0 ({\it solid line}), Z0-TURB1 ({\it dotted line}), Z0-TURB2 ({\it
    dashed line}), and Z-3-TURB2 ({\it dot-dashed line}). Note that the lines for runs Z0-TURB2 and Z-3-TURB2 overlap completely up to $t=50\; {\rm Myr}$. {\it Right
    panel:} Mass over density threshold n$_{\rm thres} = 10^3\,{\rm
    cm}^{-3}$. \label{fig5}
}
\end{figure}

\begin{figure}
\includegraphics[width=8cm]{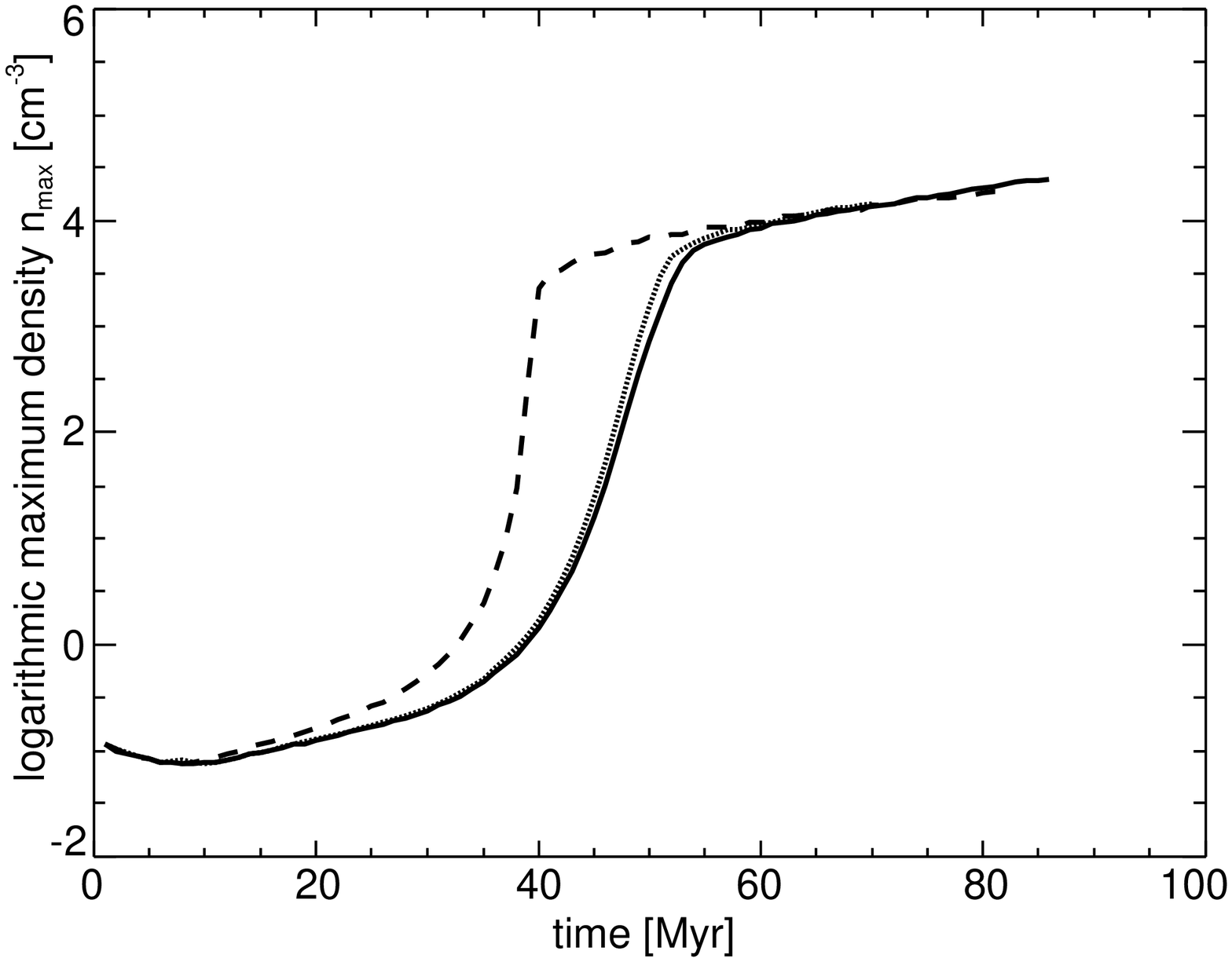}
\includegraphics[width=8.2cm]{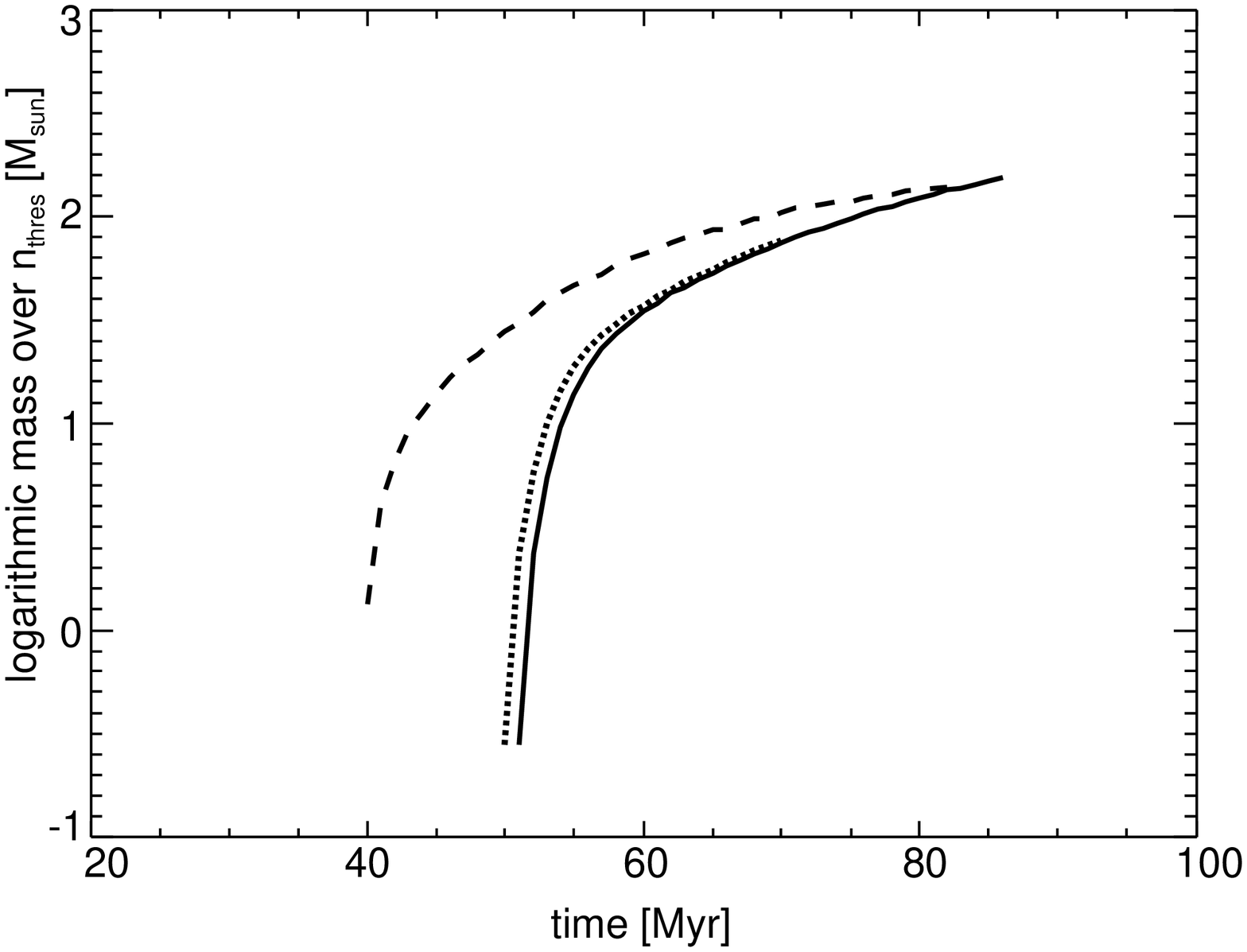}
\caption{{\it Left panel:} Maximum gas number density vs time for runs
  Z0-ROT ({\it solid line}), Z-3-ROT ({\it dotted line}), and Z-1-ROT
  ({\it dashed line}). {\it Right panel:} Mass over density threshold
  n$_{\rm thres} = 10^3\,{\rm cm}^{-3}$. \label{fig6}
}
\end{figure}

In the centrally condensed potential that we study here, a wide
variety of initial conditions for the gas yield a monolithic central collapse.
Figures~\ref{fig5} and~\ref{fig6} show that the central density depends
on the initial conditions.  Turbulence slows the growth of the central
density slightly (Fig.~\ref{fig5}), but both spherically symmetric and
turbulent initial conditions collapse and form a sink particle.  The
mass in dense gas (over an arbitrary value of $n_{\rm thres} =
10^3$~cm$^{-3}$) grows rapidly in both cases once collapse sets in
(Fig.~\ref{fig5}).

Rotation allows for the formation of stable disks over large
fractions of the local Hubble time. Figure~\ref{fig6} shows that once
a rotationally supported structure has formed, the central density and
amount of mass at high density ceases to grow quickly.  This is
because the continued growth of the central density now depends on
angular momentum transport within the rotating structure.  Molecular
viscosity is sufficiently low that no further accretion would happen
at all if that were the only cause of transport.  In modern star
formation, the additional transport in circumstellar disks occurs
because of magnetorotational instability \citep{balbus98} or
gravitational instability \citep{gammie01}.  However, our models
include no magnetic fields, so the magnetorotational instability
cannot act here.

\subsection{Gravitational Instability}

\begin{figure}
\includegraphics[width=12cm]{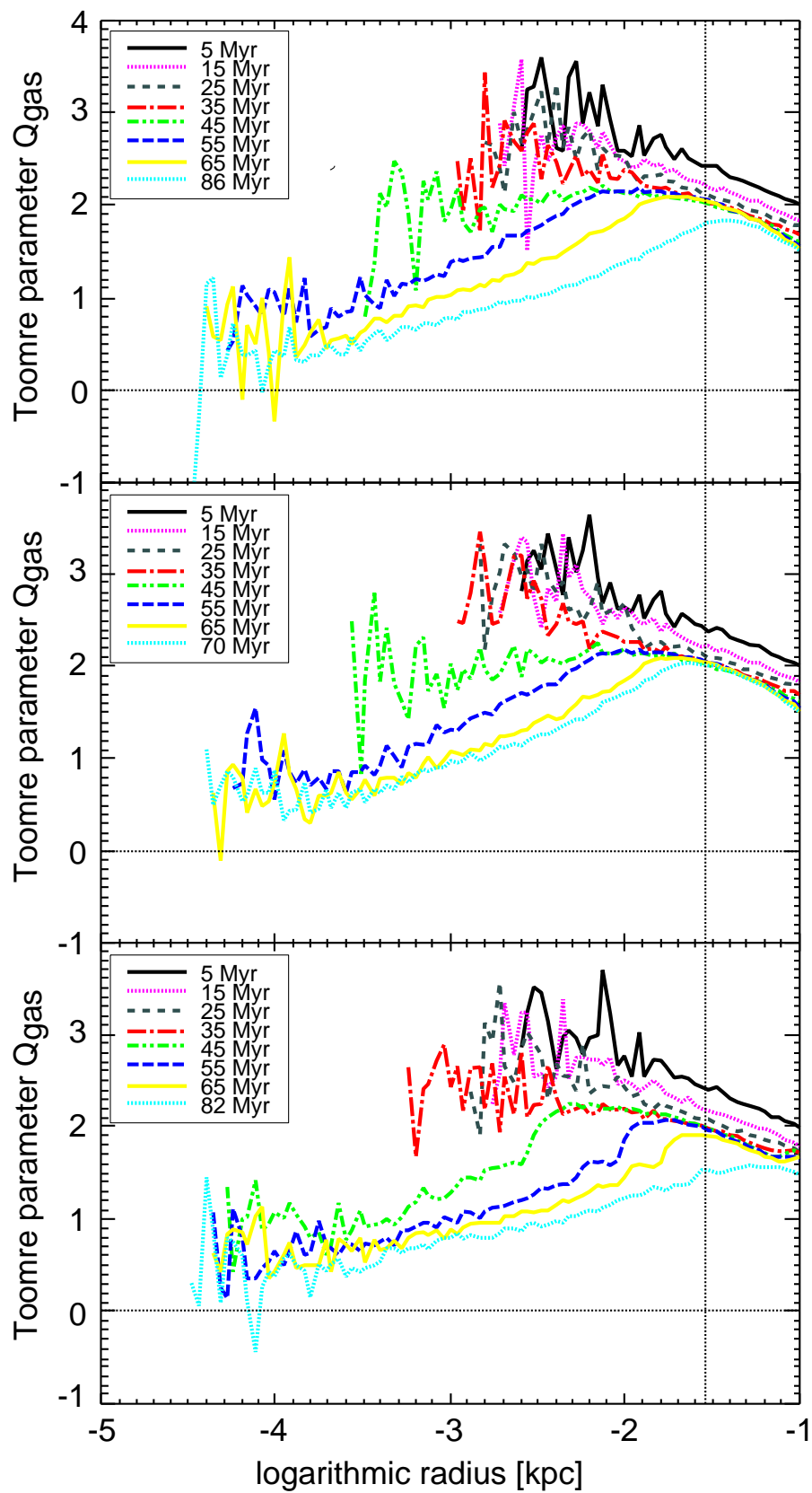}
\caption{{\it From top to bottom}: Radial distribution of the Toomre
  parameter Q for the runs with rotation Z0-ROT, Z-3-ROT, and
  Z-1-ROT. The value of $Q$ is shown on a $\log_{10}$ scale.
\label{fig7}
}
\end{figure}

To examine whether gravitational instability may be important in
driving accretion or fragmentation, we examine the value of the \citet{toomre64}
instability parameter
\begin{equation}
Q = \kappa c_{\mathrm{s}} / \pi G \Sigma,
\end{equation}
where $G$ is Newton's gravitation constant, $\Sigma$ is the disk
surface density, $\kappa$ is the epicyclic frequency, and $c_{\mathrm{s}}$ is the
sound speed. Radial gravitational instability occurs for $Q < 1$.  In
Figure~\ref{fig7} we examine our rotating runs.  We see that, in
agreement with the observed lack of fragmentation, the disks are
strongly Toomre stable for the first 50~Myr, with $Q > 10$ throughout,
and only begin to reach the unstable regime at the very center
thereafter.

In fact, given the lack of gravitational instability or magnetic
fields, the accretion towards the center seen in the simulation still
represents only an upper limit on the actual accretion rate, as all
the angular momentum transfer occurring must be due to numerical
viscosity.  In reality, virtually all galaxies interact with other
galaxies over the course of a Hubble time. The resulting tidal
perturbations drive strong accretion and subsequent star
formation.  Our results suggest that, at least at early times when
only small halos are collapsing, galaxy interactions may be essential
for driving star formation.

\begin{figure}
\epsscale{0.80}
\plotone{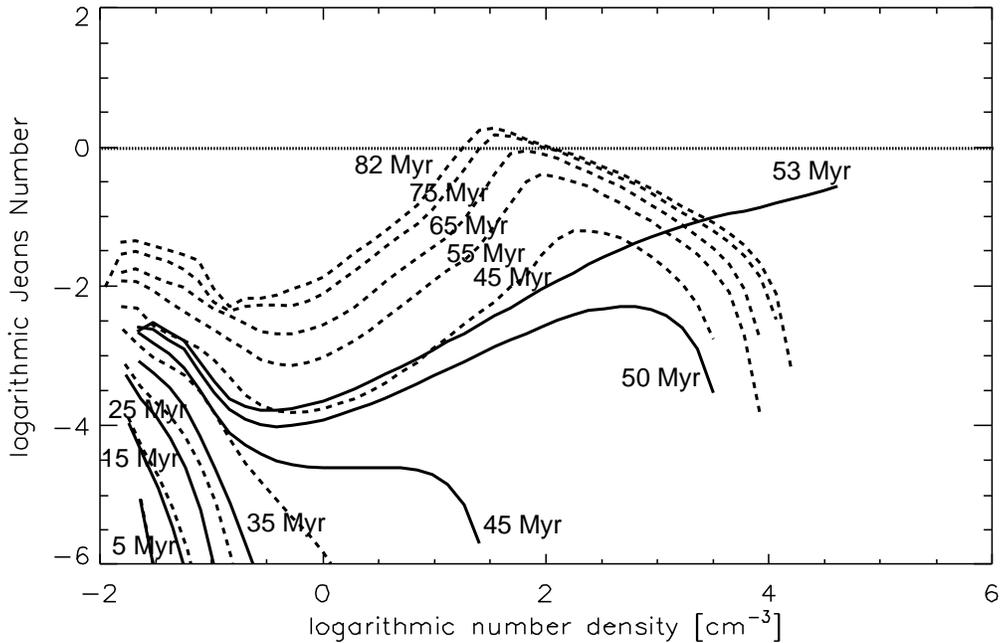}
\caption{We show the temporal evolution of the number of Jeans masses
  as a function of number density for the runs Z0 ({\it solid lines})
  and Z-1-ROT ({\it dashed lines}). The runs Z-1-ROT reach a Jeans
  number of 1 ({\it dot-dashed horizontal line}) but not at the high
  number densities that are required for efficient cooling and
  collapse. In contrast the runs Z0 show a clear trend of the Jeans
  number rising with time at high densities. The data are equally spaced in logarithmic number density with a bin size of 0.14. \label{fig8}  
}
\end{figure}

Another way of looking at the development of gravitational instability
is to examine the total number of Jeans masses at every number
density. We space our data equally in logarithmic number density with a bin size of 0.14 and in each bin we calculate the Jeans mass
\begin{equation}
M_{\mathrm{J}} =  \frac{4\pi}{3}\rho_0\left(\frac{\lambda_{\mathrm{J}}}{2}\right)^3=\frac{\pi^{5/2}}{6}G^{-3/2}\rho_0^{-1/2}{c_{\mathrm{s}}}^3, 
\end{equation}
where we define the Jeans mass $M_{\mathrm{J}}$ as the mass originally contained within a sphere of diameter $\lambda_{\mathrm{J}}$ and uniform density $\rho_0$. $\lambda_{\mathrm{J}}$ denotes the critical wavelength for the Jeans instability. Since the sound speed $c_{\mathrm{s}}$ depends on the temperature and chemical composition, we use the mean value of $c_{\mathrm{s}}$ in each density bin. We calculate the number of Jeans masses $N_{\mathrm{J}}$ above a certain density $n_0$ as
\begin{equation}
N_{\mathrm{J}} = M(n>n_0)/M_{\mathrm{J}}(n_0).
\end{equation}   
We show the result in Figure~\ref{fig8}. This compares the standard
primordial run Z0 to the highest metallicity rotating run Z-1-ROT.
Whereas Z0 reaches Jeans number unity at the highest densities at $t >
55$~Myr, Z-1-ROT only approaches Jeans number unity at lower densities
$n \sim 100$~cm$^{-3}$.  This gas is widely distributed away from the
center of the disk, however, so it cannot collect and collapse towards
a common center.  Too little material reaches the high densities at
the very center of the disk to start runaway collapse there.

\subsection{Metallicity Thresholds}

How does our result agree with other work on low-metallicity star
formation? We have already discussed the comparison with
\citet{bcl01}.  The same authors examined their zero metallicity case
in the absence of an ambient UV radiation sufficient to dissociate
H$_2$, so that molecules can form and contribute to the cooling.  They
found the same result as was later reached by Paper IV, that the top
hat potential does lead to fragmentation even at zero metallicity
\citep{bcl02}.

A more advanced computation was performed by \citet{ss07}, who used the
adaptive mesh grid code Enzo \citep{o04} and a comprehensive treatment
of metal cooling rates as detailed by \citet{ssa08}. Instead of a
top-hat dark matter potential, they used cosmological initial
conditions, with collapse followed over 28 levels of refinement.  Like
\citet{bcl02} they assumed no dissociating radiation and full
molecular cooling.  As a result of their choice of initial conditions,
their collapsing objects were centrally concentrated, similar to
ours.  In agreement with our results, they found no fragmentation
below densities of $10^5$~cm$^{-3}$.  At higher densities of
$10^8$~cm$^{-3}$, they did see hints of fragmentation of their central
object into two to four objects at metallicities $Z >
10^{-4}\,Z_{\odot}$, but that alone seems insufficient to mark a shift
from a primordial to a modern initial mass function.

Following up on that work \citet{SMI08} presented a suite of simulations investigating the critical metallicity threshold. Two of their runs show fragmentation below densities of $10^5$~cm$^{-3}$ into 2 bound clumps but most of their runs show fragmentation at higher densities. Their set of models which collapses at a redshift similar to the one used in our calculations only results in 2 fragments at a metallicity of $Z = 10^{-3.5}\,Z_{\odot}$ and $Z = 10^{-2}\,Z_{\odot}$. These differences could also be due to the different initial conditions used. Our simulations began with hot, ionized gas, whereas their simulations started with cold neutral gas. In Paper II we explained the consequences of these differences in more detail.
In agreement with \citet{SMI08} we also find that the gas reaching the CMB temperature results in a thermally stable gas cloud.

\citet{sch02,sch06} and \citet{omu05} propose that rather than atomic
or molecular line cooling determining the metallicity where strong
fragmentation sets in, dust continuum cooling does so. Surprisingly,
their estimates of cooling rates suggest that fragmentation at high
densities $n > 10^{10}$~cm$^{-3}$ already sets in for metallicities as
low as $Z > 10^{-6}\,Z_{\odot}$. \citet{omu05} demonstrates that even
at these low metallicities, dust cooling gives an effective adiabatic
index under unity for a range of densities $10^{10} < n <
10^{14}$~cm$^{-3}$, promoting rapid fragmentation of objects
collapsing through that density range.

High resolution nested-grid models by \citet{momi08}
show binaries forming even at primordial metallicity $Z= 0$ at
densities $ n > 10^{16}$\ cm$^{-3}$. \citet{m08} then found that
binary fragmentation is actually enhanced at decreasing metallicity,
with the rotation threshold for fragmentation at $Z=0$ lying a full two orders
of magnitude lower than for solar metallicity $Z=Z_{\odot}$. Low metallicity
fragmentation occurs at densities as high as $10^{16}$~cm$^{-3}$
for $Z = 0$, in contrast to modern star formation where binaries only
form for $10^{11} < n < 10^{15}$~cm$^{-3}$.  However, primordial
binary formation still yields very massive stars, simply in pairs
rather than alone.

Further simulation of collapsing cores using the \citet{omu05}
equation of state has shown that fragmentation promoted by dust
cooling at high densities indeed appears able to shift star formation
from a primordial to a more present-day-like initial mass function.
\citet{to06} followed fragmentation of low metallicity $Z <
10^{-4}\,Z_{\odot}$, high density $n > 10^{10}$~cm$^{-3}$ gas in the
non-rotating case. \citet{clark08} included rotation, and used sink
particles to follow the collapse of a complete cluster.  They agree
with \citet{momi08} in finding only small numbers of high-mass
fragments in the primordial case, but already at metallicities of $Z
\sim 10^{-5}\,Z_{\odot}$, they find an average mass under $1\,M_{\odot}$,
approaching a modern initial mass function. 

Dust cooling looks likely to be the primary physics determining the
shift in typical stellar mass from primordial to modern stars, with a
threshold of $Z \sim 10^{-5}\,Z_{\odot}$ consistent with the lowest
metallicity stars observed in the modern galaxy.  Conversely, as we
have shown in Paper~IV and this paper, atomic line cooling does not
appear to dominate this transition, and it appears unlikely that the
transition occurs at metallicities as high as $Z =
10^{-3}\,Z_{\odot}$.

\acknowledgments This research was supported in part by the National
Science Foundation under Grant No. PHY05-51164. AKJ acknowledges
support by the Human Resources and Mobility Programme of the European
Community under the contract MEIF-CT-2006-039569. M-MML was partly
supported by stipends from the Max-Planck-Gesellschaft and the
Deutsche Akademische Austausch Dienst, and acknowledges the
hospitality of the Waldkindergarten Heidelberg during the initial
drafting of this paper. RSK acknowledges support from
the German science foundation under the Emmy Noether grant
KL1358/1 and the Priority Program SFB 439 {\em Galaxies in
the Early Universe}. SCOG acknowledges funding from the
Germany science foundation via grant KL1358/4. SK kindly acknowledges support by an EU Commission "Marie Curie Intra-European (Individual) Fellowship" of the 6th Framework Programme with contract number MEIF-CT-2004-011226. Computations were performed at the McKenzie cluster at the Canadian Institute for Theoretical Astrophysics and the Sanssouci cluster at the Astrophysikalisches Institut Potsdam. Figures~\ref{fig3} and \ref{fig4} were produced using SPLASH, a visualization package for SPH written by \citet{pri07}.

\appendix
\section{Particle splitting}
\label{appendix}
In Paper II we presented low-resolution simulations that showed that
the density increases most rapidly close to the center of the dark
matter halo. On-the-fly splitting \citep{KIT02} with two levels of
refinement at 2 different radii provides us with the highest mass
resolution at the region of interest. Every SPH particle that crosses
a radius~$r_1$ of 0.3~kpc towards smaller radii acts as a parent
particle and spawns 13 child particles with a mass of
$m_{\mathrm{p}}/13$. The same happens at a radius~$r_2$ of 0.2 kpc
where gas particles crossing $r_2$ are split again. The mass of the
gas particles within $r_2$ is thus $m_{\mathrm{p}}/169 \approx
1.5\times 10^{-2}\,M_{\odot}$. In the region of collapse we can thus
resolve Jeans masses down to $1.5\,M_{\odot}$, as compared to
$M_{\mathrm{res}} = 200\,M_{\odot}$ in the low resolution simulations.
Each child particle inherits directly velocity, internal energy and
fractional abundances of the chemical species. Subsequently the child
particles are evolved with standard SPH procedures. The only
difference is that, to mitigate interactions between adjacent
particles having different masses, we have modified the scheme by
which GADGET calculates the smoothing lengths of particles. We now
evolve $h_i$, the smoothing length for particle $i$, so that the
radius $h_i$ contains between $N_{\mathrm{neigh}}-N_{\mathrm{dev}}$
and $N_{\mathrm{neigh}}+N_{\mathrm{dev}}$, where $N_{\mathrm{dev}}$ is
the allowed deviation in the number of neighbors, as well as a mass
between $N_{\mathrm{neigh}}-N_{\mathrm{dev}}$ and
$N_{\mathrm{neigh}}+N_{\mathrm{dev}}$ times the mass of particle
$i$. If both conditions cannot be satisfied simultaneously we allow
for a larger number of neighbors or a larger amount of mass within a
sphere of radius $h_i$.  For one specific set of parameters we carried
out two simulations, one with the refinement technique and the other
without it, and verified that temporal perturbations caused by
refinement do not affect the results presented in Section
\ref{sec:result}.

\clearpage

\clearpage
\clearpage
\clearpage

\clearpage

\begin{deluxetable}{cl}
\tablewidth{0pt}
\tablecaption{\label{tab:species}}
\tablehead{Species & Method of solution}
\startdata
$\me$ & Conservation law \\
$\Hp$ & Rate equation \\
$\mH$ & Conservation law \\
$\Hm$ & Equilibrium abundance \\
$\mHtp$ & Equilibrium abundance \\
$\mHt$ & Rate equation \\
$\htp$ & Equilibrium abundance \\
$\Dp$ & Rate equation \\
$\mD$ & Conservation law \\
$\hd$ & Rate equation \\
$\Hep$ & Rate equation \\
$\He$ & Conservation law \\
$\Cp$ & Rate equation \\
$\mC$ & Conservation law \\
$\Cm$ & Equilibrium abundance \\
$\Op$ & Rate equation \\
$\mO$ & Conservation law \\
$\Om$ & Equilibrium abundance \\
$\sip$ & Rate equation \\
$\msi$ & Conservation law \\
$\sipp$ & Rate equation \\
$\ch^{+}$ & Equilibrium abundance \\
$\ch_{2}^{+}$ & Equilibrium abundance \\
$\ch_{3}^{+}$ & Rate equation \\
$\ch$ & Rate equation \\
$\ch_{2}$ & Rate equation \\
$\oh^{+}$ & Equilibrium abundance \\
$\hto^{+}$ & Equilibrium abundance \\
${\rm H_{3}O^{+}}$ & Equilibrium abundance \\
$\oh$ & Rate equation \\
$\hto$ & Rate equation \\
$\co^{+}$ & Equilibrium abundance \\
$\co$ & Rate equation \\
${\rm HOC^{+}}$ & Equilibrium abundance \\
${\rm HCO^{+}}$ & Rate equation \\
${\rm O_{2}^{+}}$ & Equilibrium abundance \\
${\rm O_{2}}$ & Rate equation \\
${\rm C_{2}}$ & Rate equation \\
${\rm SiH^{+}}$ & Equilibrium abundance \\
\enddata
\end{deluxetable}

\begin{deluxetable}{lc}
\tabletypesize{\small}
\tablecaption{Processes included in our thermal model. \label{cool_model}}
\tablewidth{0pt}
\tablehead{
\colhead{Process}  & \colhead{References} }
\startdata
{\bf Cooling:} & \\
Lyman-$\alpha$ cooling & \citet{CEN92} \\
He electronic excitation & \citet{CEN92,BBFT00} \\
Thermal bremsstrahlung & \citet{sk87} \\
Compton cooling & \citet{CEN92} \\
$\mHt$ rotational, vibrational lines & \citet{BOU99,ra04} \\
$\hd$ rotational, vibrational lines  & \citet{lna05} \\
Fine structure lines ($\mC, \Cp, \mO, \msi, \sip$) & Many sources; see papers I, III \\
CO rotational, vibrational lines & \citet{nk93,nlm95} \\
${\rm H_{2}O}$ rotational, vibrational lines & \citet{nk93,nlm95} \\
OH rotational, vibrational lines  lines & \citet{pav02} \\
$\Hp$ recombination & \citet{FER92} \\
$\Hep$ recombination & \citet{ap73,hs98} \\
$\mH$ \& $\He$ collisional ionization & \citet{JAN87} \\
$\mHt$ collisional dissociation & Many sources; see papers I, III \\
{\bf Heating:} & \\
$\mHt$ gas-phase formation & Many sources; see papers I, III \\
\enddata
\end{deluxetable}

\begin{deluxetable}{llllrrcc}
\tablewidth{0pt}
\tablecaption{Physical state of the densest gas within the scale radius $r_{\mathrm{s}}$
at time $t_{\rm end}$ \label{tab:runs}}
\tablehead{\colhead{Run} & \colhead{$Z$\tablenotemark{a}} & \colhead{$\lambda$\tablenotemark{b}} & \colhead{$E_{\rm turb}$\tablenotemark{c}} &
\colhead{$t_{\rm end}$\tablenotemark{d}} &\colhead{$T_{\rm c, min}$\tablenotemark{e}} &
\colhead{$n_{\rm c, max}$\tablenotemark{f}} &
\colhead{$x_{\rm \mHt, c, max}$\tablenotemark{g}}
\\\colhead{} &\colhead{($Z_{\odot}$)} & \colhead{} & \colhead{($E_{\rm int}$)} &  \colhead{(Myrs)} & \colhead{(K)} &
\colhead{(${\rm cm^{-3}}$)} & \colhead{}}
\startdata
Z0 & 0.0 & 0.0 & 0.0 & 53 & 79 & $> n_{\rm sink}$ & $2.9 \times 10^{-3}$\\
Z0-TURB1 & 0.0 & 0.0 & 0.05 & 71 & 66 & $> n_{\rm sink}$ & $3.2 \times 10^{-3}$\\
Z0-TURB2 & 0.0 & 0.0 & 0.1 & 50 & 82 & $3.5 \times 10^{2}$ & $2.8 \times 10^{-3}$\\
Z-3-TURB2 & $10^{-3}$ & 0.0 & 0.1 & 65 & 66 & $1.6 \times 10^{4}$ & $3.0 \times 10^{-3}$\\
Z0-ROT & 0.0 & 0.05 & 0.0 & 86 & 67 & $2.5 \times 10^{4}$ & $2.9 \times 10^{-3}$\\
Z-3-ROT & $10^{-3}$ & 0.05 & 0.0 & 70 & 59 & $1.5 \times 10^{4}$ & $2.9 \times 10^{-3}$\\
Z-1-ROT & $10^{-1}$ & 0.05 & 0.0 & 82 & 67 & $1.9 \times 10^{4}$ & $2.7 \times 10^{-3}$\\
\enddata
\tablenotetext{a}{Metallicity of the gas.}
\tablenotetext{b}{Spin parameter of the halo.}
\tablenotetext{c}{Added turbulent energy in units of the initial internal energy.}
\tablenotetext{d}{Time at the end of the simulation.}
\tablenotetext{e}{Minimum temperature of the gas within the scale radius~$r_{\rm s}$.}
\tablenotetext{f}{Maximum number density of the gas within the scale radius~$r_{\rm s}$. $n_{\rm sink}=1.25 \times 10^5\,{\rm cm^{-3}}$ is the threshold density for sink creation.}
\tablenotetext{g}{Maximum fractional H$_2$ abundance within the scale radius~$r_{\rm s}$.}
\end{deluxetable}


\begin{thebibliography}{}
\bibitem[Abel, Bryan, \& Norman(2002)]{abn02}
Abel, T., Bryan, G.~L., \& Norman, M.~L. 2002, Science, 295, 93

\bibitem[Aldrovandi \& Pequignot(1973)]{ap73}
Aldrovandi, S.~M.~V. \& Pequignot, D. 1973, A\&A, 25, 137

\bibitem[Balbus \& Hawley(1998)]{balbus98} Balbus, S. A., \& Hawley,
  J. F. 1998, Rev.\ Mod.\ Phys., 70, 1

\bibitem[{{Bate} {et~al.}(1995){Bate}, {Bonnell}, \& {Price}}]{BAT95}
{Bate}, M.~R., {Bonnell}, I.~A., \& {Price}, N.~M. 1995, \mnras, 277,
362

\bibitem[{{Bate} \& {Burkert}(1997)}]{BAT97}
{Bate}, M.~R. \& {Burkert}, A. 1997, \mnras, 288, 1060

\bibitem[Beers \& Christlieb(2005)]{bc05}
Beers, T.~C., \& Christlieb, N. 2005, ARA\&A, 43, 531

\bibitem[{{Benz}(1990)}]{ben90}
{Benz}, W. 1990, in Numerical Modelling of Nonlinear Stellar Pulsations
Problems and Prospects, ed. J.~R. Buchler (Dordrecht: Kluwer), 269

\bibitem[Bray et~al.(2000)]{BBFT00}
Bray, I., Burgess, A., Fursa, D.~V., \& Tully, J.~A. 2000, A\&AS, 146, 481

\bibitem[Bromm et~al.(2001)]{bcl01}
{Bromm}, V., {Ferrara}, A., {Coppi}, P.~S., \& {Larson}, R.~B. 2001, \mnras, 328, 969

\bibitem[{{Bromm} {et~al.}(2002){Bromm}, {Coppi}, \& {Larson}}]{bcl02}
{Bromm}, V., {Coppi}, P.~S., \& {Larson}, R.~B. 2002, \apj, 564, 23

\bibitem[Brown, Byrne, \& Hindmarsh(1989)]{bbh89}
Brown, P.~N., Byrne, G.~D., \& Hindmarsh, A.~C. 1989,
SIAM J.\ Sci.\ Stat.\ Comput., 10, 1038

\bibitem[{{Cen}(1992)}]{CEN92}
{Cen}, R. 1992, \apjs, 78, 341

\bibitem[Clark et al.(2008)]{clark08} Clark, P. C., Glover, S. C. O., \&
  Klessen, R. S. 2008, \apj, 672, 757

\bibitem[Einasto(1965)]{einasto65} Einasto, J. 1965, Trudy Inst.\
  Astrofiz.\ Alma-Ata, 5, 87

\bibitem[Ferland et~al.(1992)]{FER92}
Ferland, G.~J., Peterson, B.~M., Horne, K., Welsh, W.~F., \& Nahar, S.~N. 1992, \apj, 387, 95

\bibitem[Frebel, Johnson \& Bromm(2007)]{fjb07}
Frebel, A., Johnson, J.~L., \& Bromm, V. 2007, \mnras, 380, 40 

\bibitem[Gammie(2001)]{gammie01} Gammie, C. F. 2001, \apj, 553, 174

\bibitem[Gao et al.(2008)]{gao08} Gao, L., Navarro, J. F., Cole, S.,
  Frenk, C., White, S. D. M., Springel, V., Jenkins, A., Neto,
  A. F. 2008,  \mnras, 387, 536

\bibitem[Glover(2008)]{glo07}
Glover, S.~C.~O. 2008, in preparation. (Paper III) 

\bibitem[Glover \& Jappsen(2007)]{gj07}
Glover, S.~C.~O., \& Jappsen, A.-K. 2007, \apj, 666, 1 (Paper I)

\bibitem[Hummer \& Storey(1998)]{hs98}
Hummer, D.~G., \& Storey, P.J. 1998, \mnras, 297, 1073

\bibitem[Janev et~al.(1987)]{JAN87}
Janev, R.~K., Langer, W.~D., Evans, K., \& Post, D.~E. 1987, Elementary
Processes in Hydrogen-Helium Plasmas, Springer

\bibitem[{{Jappsen} {et~al.}(2005){Jappsen}, {Klessen}, {Larson}, {Li}, \& {Mac Low}}]{JAP05}
{Jappsen}, A.-K., {Klessen}, R.~S., {Larson}, R.~B., {Li}, Y., \& {Mac Low}, M.-M. 2005,
A\&A, 435, 611

\bibitem[Jappsen et~al.(2007a)]{jgkm07}
Jappsen, A.-K., Glover, S.~C.~O., Klessen, R.~S., \& {Mac Low}, M.-M. 2007a, \apj, 660, 1332 (Paper II)

\bibitem[Jappsen et al.(2007b)]{jkgm07}
Jappsen, A.-K., Klessen, R.~S., Glover, S.~C.~O.,
\& {Mac Low}, M.-M. 2007b, submitted to ApJ, arXiv:0709.3530 (Paper IV)

\bibitem[Kitsionas \& Whitworth(2002)]{KIT02}
Kitsionas, S. \& Whitworth, A.~P. 2002, \mnras, 330, 129

\bibitem[{{Le Bourlot} {et~al.}(1999){Le Bourlot}, {Pineau des For{\^ e}ts},
    \& {Flower}}]{BOU99}
{Le Bourlot}, J., {Pineau des For{\^ e}ts}, G., \& {Flower}, D.~R. 1999, \mnras,
305, 802

\bibitem[Lipovka, N\'u\~nez-L\'opez, \& Avila-Reese(2005)]{lna05}
Lipovka, A., N\'u\~nez-L\'opez, R., \& Avila-Reese, V. 2005, \mnras, 361, 850

\bibitem[Mac Low et al.(1998)]{MAC98}
{Mac Low}, M.-M., {Klessen}, R.~S., {Burkert}, A., {Smith}, M.~D. 1998, Phys.\ Rev.\ Lett., 80, 2754

\bibitem[{{Mac Low}(1999)}]{MAC99}
{Mac Low}, M.-M. 1999, \apj, 524, 169

\bibitem[Machida(2008)]{m08} Machida, M. N. 2008, \apj, 682, L1

\bibitem[Machida et al.(2008)]{momi08} Machida, M. N., Omukai, K.,
  Matsumoto, T., \& Inutsuka, S.-I. 2008, \apj, 677, 813

\bibitem[Matsumoto et al.(1997)]{matsumoto97} Matsumoto, T., Hanawa,
  T., \& Nakamura, F. 1997, \apj, 478, 569

\bibitem[McKee \& Tan(2008)]{mck08}
McKee, C.~F. \& Tan, J.~C., \apj, 681, 771

\bibitem[Merritt et al.(2006)]{merritt06}  Merritt,  D., Graham,  A. W.,
  Moore,  B., Diemand,  J., Terzi\'c,  B., 2006, AJ, 132, 2685

\bibitem[{{Monaghan}(1992)}]{mon92}
{Monaghan}, J.~J. 1992, ARA\&A, 30, 543

\bibitem[{{Monaghan}(2005)}]{mon05}
{Monaghan}, J.~J. 2005, Rep.\ Prog.\ Phys., 68, 1703

\bibitem[{{Navarro} {et~al.}(1997){Navarro}, {Frenk}, \& {White}}]{nfw97}
{Navarro}, J.~F., {Frenk}, C.~S., \& {White}, S.~D.~M. 1997, \apj, 490, 493

\bibitem[Navarro et al.(2004)]{navarro04} Navarro, J. F., Hayashi, E.;
Power, C., Jenkins, A. R., Frenk, C. S., White, S. D. M., Springel,
V., Stadel, J. \& Quinn, T. R. 2004, MNRAS, 349, 1039

\bibitem[Neufeld \& Kaufman(1993)]{nk93}
Neufeld, D.~A., \& Kaufman, M.~J. 1993, \apj, 418, 263

\bibitem[Neufeld, Lepp \& Melnick(1995)]{nlm95}
Neufeld, D.~A., Lepp., S., \& Melnick, G.~J. 1995, ApJS, 100, 132

\bibitem[{{Oh} \& {Haiman}(2003)}]{oh03}
{Oh}, S.~P. \& {Haiman}, Z. 2003, \mnras, 346, 456

\bibitem[Omukai \& Palla(2001)]{omu01}
Omukai, K. \& Palla, F., 2001, \apj, 561, L55

\bibitem[Omukai \& Palla(2003)]{omu03}
Omukai, K. \& Palla, F., 2003, \apj, 589, 677

\bibitem[Omukai et~al.(2005)]{omu05}
Omukai, K., Tsuribe, T., Schneider, R., \& Ferrara, A. 2005, ApJ, 626, 627

\bibitem[O'Shea et al.(2004)]{o04} O'Shea, B. W., Bryan, G., Bordner,
  J., Norman, M. L., Abel, T., Harkness, R., \& Kritsuk, A. 2004, in
  Adaptive Mesh Refinement---Theory and Applications, eds. T. Plewa,
  T. Linde, \& V. G. Weirs (Berlin, Springer), 341

\bibitem[O'Shea \& Norman(2006)]{oshn06}
O'Shea, B.~W., \& Norman, M.~L. 2006, \apj, 654, 66

\bibitem[Pavlovski et~al.(2002)]{pav02}
Pavlovski, G., Smith, M.~D., {Mac Low}, M.-M., \& Rosen, A. 2002, MNRAS,
337, 477

\bibitem[Peebles(1971)]{peeb71}
Peebles, P.~J.~E. 1971, A\&A, 11, 377

\bibitem[{{Pettini}(1999)}]{PET99}
{Pettini}, M. 1999, in Chemical Evolution from Zero to
High Redshift, eds. J.~R. Walsh, M.~R. Rosa, (Berlin: Springer), 233

\bibitem[Prada et al.(2006)]{prada06} Prada, F., Klypin, A. A.,
  Simonneau, E., Betancort-Rijo, J., Patiri, S., Gottlber, S.,
  Sanchez-Conde, M. A. 2006, \apj, 645, 1001

\bibitem[Price(2007)]{pri07}
Price, D.~J. 2007, PASA, 24, 159 

\bibitem[Ricotti \& Ostriker(2004)]{RIC04}
Ricotti, M. \& Ostriker, J.~P. 2004, \mnras, 350, 539

\bibitem[Ripamonti \& Abel(2004)]{ra04}
Ripamonti, E., \& Abel, T. 2004, \mnras, 348, 1019

\bibitem[Schneider et al.(2002)]{sch02} Schneider, R., Ferrara, A.,
  Natarajan, P., \& Omukai, K. 2002, \apj, 571, 30

\bibitem[Schneider et al.(2006)]{sch06} Schneider, R., Omukai, K.,
  Inoue, A. K., \& Ferrara, A. 2006, \mnras, 369, 1437

\bibitem[Shapiro \& Kang(1987)]{sk87}
Shapiro, P.~R., \& Kang, H. 1987, ApJ,  318, 32

\bibitem[Smith \& Sigurdsson(2007)]{ss07} Smith, B. D., \& Sigurdsson,
  S. 2007, \apj, 661, L5

\bibitem[Smith et al.(2008a)]{ssa08} Smith, B. D., Sigurdsson, S., \&
  Abel, T. 2008, \mnras, 385, 1443

\bibitem[Smith et al.(2008b)]{SMI08} Smith, B. D., Turk, M. J., Sigurdsson, S., O'Shea, B. W. \& Norman, M. L. 2008, accepted by \apj, arXiv:0806.1653v2 

\bibitem[{{Spergel} {et~al.}(2003)}]{spe03}
{Spergel}, D.~N., {Verde}, L., {Peiris}, H.~V., {et~al.} 2003, \apjs,
148, 175

\bibitem[{{Springel} {et~al.}(2001){Springel}, {Yoshida}, \& {White}}]{syw01}
{Springel}, V., {Yoshida}, N., \& {White}, S.~D.~M. 2001, New Astron., 6, 79

\bibitem[Springel(2005)]{spr05}
Springel, V. 2005, MNRAS, 364, 1105

\bibitem[Tominaga, Umeda \& Nomoto(2007)]{tun07}
Tominaga, N., Umeda, H., \& Nomoto, K. 2007, \apj, 660, 516

\bibitem[Toomre(1964)]{toomre64} Toomre, J. 1964, \apj, 139, 1217

\bibitem[Tsuribe \& Omukai(2006)]{to06} Tsuribe, T., \& Omukai,
  K. 2006, \apj, 642, L61

\bibitem[{{Yoshida} {et~al.}(2003){Yoshida}, {Abel}, {Hernquist}, \& {Sugiyama}}]{YOS03}

{Yoshida}, N., {Abel}, T., {Hernquist}, L., \& {Sugiyama}, N. 2003, \apj, 592,
645

\bibitem[Yoshida {\em et~al.}(2006)]{yoha06}
Yoshida, N., Omukai, K., Hernquist, L., \& Abel, T. 2006, ApJ, 652, 6

\end{thebibliography}
\end{document}